\documentclass[useAMS,usenatbib]{mn2e}
\usepackage{graphicx}
\usepackage[fleqn]{amsmath}
\usepackage{amssymb}
\usepackage{multirow}

\def\ie{i.e.~}
\def\eg{e.g.~}
\def\HST{{\it{HST}}}

\def\D{\mathcal{D}}
\def\S{\mathcal{S}}
\def\N{\mathcal{N}}

\def\th{\bmath\theta}
\def\ph{\bmath\phi}
\def\bz{\bmath0}
\def\bsig{\bmath\sigma}
\def\Bsig{\bmath\Sigma}
\def\I{\mathbfss{I}}
\def\bt{\bmath t}
\def\bflux{\bmath f}
\def\bx{\bmath x}

\def\E{\mathbb{E}}
\def\bep{\bmath \epsilon}
\def\bmu{\bmath \mu}
\def\0{\mathbf{0}}
\def\AICC{AIC$_\mathrm{C}$}

\title[Reliable inference of light curve parameters]{Reliable inference of exoplanet light curve parameters using deterministic and stochastic systematics models}
\author[N. P. Gibson]{
N. P. Gibson$^{1}$\thanks{E-mail: ngibson@eso.org}
\smallskip
\\
$^{1}$European Southern Observatory, Karl-Schwarzschild-Str. 2, 85748 Garching bei M\"unchen, Germany\\
}

\begin{document}


\pagerange{\pageref{firstpage}--\pageref{lastpage}} \pubyear{2002}

\maketitle

\label{firstpage}

\begin{abstract}

Time-series photometry and spectroscopy of transiting exoplanets allow us to study their atmospheres. Unfortunately, the required precision to extract atmospheric information surpasses the design specifications of most general purpose instrumentation, resulting in instrumental systematics in the light curves that are typically larger than the target precision. Systematics must therefore be modelled, leaving the inference of light curve parameters conditioned on the subjective choice of systematics models and model selection criteria. This paper aims to test the reliability of the most commonly used deterministic systematics models and model selection criteria. As we are primarily interested in recovering light curve parameters rather than the favoured systematics model, marginalisation over systematics models is introduced as a more robust alternative than simple model selection. This can incorporate uncertainties in the choice of systematics model into the error budget as well as the model parameters. Its use is demonstrated using a series of simulated transit light curves. Stochastic models, specifically Gaussian processes, are also discussed in the context of marginalisation over systematics models, and are found to reliably recover the transit parameters for a wide range of systematics functions. None of the tested model selection criteria -- including the Bayesian Information Criterion -- routinely recovered the correct model. This means that commonly used methods that are based on simple model selection may underestimate the uncertainties when extracting transmission and eclipse spectra from real data, and low-significance claims using such techniques should be treated with caution. In general, no systematics modelling techniques are perfect; however, marginalisation over many systematics models helps to mitigate poor model selection, and stochastic processes provide an even more flexible approach to modelling instrumental systematics.

\end{abstract}


\section{Introduction}

Transiting exoplanets allow us to study their compositions and atmospheres via wavelength dependent variations in their light curves, and significant progress has been made in this field since the first detection of a exo-planetary atmosphere \citep{Charbonneau_2002}. During primary transit we can extract a transmission spectrum: a measurement of the transit depth or effective planet radius as a function of wavelength, which in turn depends on opacities in the planet's upper atmosphere and therefore can reveal the presence of atomic and molecular species \citep{Seager_2000,Brown_2001}. During an eclipse or secondary transit, when the planet passes behind its host star, the eclipse depth is a measure of the star-planet flux ratio \citep[\eg][]{Charbonneau_2005,Deming_2005}, and measuring this at multiple wavelengths allows us to construct a spectrum of the planet's dayside. Finally, phase curves can reveal the day/night contrast and map a planet's emission spectrum as a function of longitude \citep[\eg][]{Knutson_2007,Cowan_2008,Knutson_2009}.

All of these methods require exquisite temporal stability in the light curves in order to measure the tiny wavelength dependent changes that the planetary atmosphere imprints on the light curves (typically $\sim10^{-4}$ for hot-Jupiters). In terms of photon noise, this is easily achievable for nearby, bright stellar systems. However, there are currently no dedicated instruments to study exoplanet atmospheres, and consequently we have been using common-user instruments for many years that were not designed with this level of precision in mind. This means that the vast majority of our observations are limited by our understanding of the instrumental systematics, which manifest themselves as bumps and wiggles in the light curves, often of unknown origin. These are usually larger than the target precision of our observations, and must therefore be modelled when extracting measurements from exoplanet light curves in order to reach a useful precision.

Despite considerable efforts, systematics models remain a key factor in the analysis and interpretation of exoplanet spectra, and the community is yet to reach a consensus on the correct approaches to obtaining robust results in the presence of systematic noise. One prominent example is the NICMOS transmission spectrum of the hot Jupiter HD 189733b. The original analysis was performed using a simple, linear basis model to account for the systematics, where the bases consist of auxiliary parameters related to the orbital phase and telescope pointing. This resulted in the detection of molecular species \citep{Swain_2008}, but the choice of linear basis model was arbitrary rather than physically motivated.
\citet{Gibson_2011} showed that this method was particularly unreliable for NICMOS, and the detection of molecular species was therefore inconclusive. Since then, subsequent analyses of the same data set have revealed significantly larger uncertainties using more objective models such as Gaussian processes \citep{Gibson_2012} and Independent Component Analysis \citep{Waldmann_2013}. Furthermore, reanalyses of other NICMOS data sets have shown the linear basis model approach to produce inconsistent results \citep{Crouzet_2012}, and new observations of the same targets with WFC3 have also showed that the NICMOS results are unreliable \citep{Deming_2013}. However, in spite of this mounting evidence, the simple, arbitrary basis model approach is still used for inferring molecular features in transmission and eclipse spectra \citep{Swain_2014}.

Indeed, such deterministic systematics models based on linear basis functions remain the default technique for extracting light curve parameters, in part because they are easy to implement, fast to compute and have a simple interpretation.
Most studies have progressed to using model selection criteria to discriminate between a wide selection of models resulting in more objective inference \citep[\eg][]{Sing_2011,Huitson_2013,Nikolov_2014,Crossfield_2013}. Nonetheless, much of our knowledge of exoplanet atmospheres is based on this approach, and there is yet to be a systematic study investigating its validity. This paper aims to rectify this, by placing some of the commonly used systematics models in context, and testing our current methods of model selection in recovering the `true' systematics model. New techniques are introduced and tested alongside them, based on marginalising over multiple systematics models rather than selecting one, and the use of shrinkage priors for model selection and to mitigate the risk of over-fitting. These tests are designed to inform on the best practices for modelling instrumental systematics when inferring light curve parameters.

The paper is organised as follows: Sect.~\ref{sect:inference} introduces the basic modelling of exoplanet light curves and the principles of model selection. Sect.~\ref{sect:model_marg} introduces marginalising over multiple systematics rather than simple model selection, describes how a Laplacian approximation can be used to calculate the Bayesian evidence, and discusses Gaussian processes in the context of marginalisation over systematics models. Sect.~\ref{sect:simulations} describes a series of tests of these systematics models using simulated transit light curves. Finally Sect.~\ref{sect:practical} discusses the best practices for model selection based on the results, and Sect.~\ref{sect:con} concludes.

\section{Inference of light curve parameters}
\label{sect:inference}

To discuss the impact of systematics models on the inference of light curve parameters, it is convenient to begin with a description of a light curve model. The light curve consists of a series of $N$ flux measurements given by the column vector $\bflux = (f_{1} ,\ldots, f_{N})^T$, observed at time $\bt = (t_{1} ,\ldots, t_{N})^T$, collectively referred to as the data, $\D$. We fit a model light curve $T(t,\th)$, a function of time with model parameters $\th$, and wish to infer the joint probability distribution of $\th$.

Under the assumption that the flux measurements are Gaussian distributed about the light curve model with mean zero and standard deviation $\bsig = (\sigma_{1} ,\ldots, \sigma_{N})^T$, we proceed by constructing a likelihood function as the product of independent Gaussians:
\begin{align*}
 p(\D | \th) &= \prod_{i=1}^N \N (f_i | T(t_i,\th), \sigma_i^2) \\ 
 &= \N (\bflux | T(\bt,\th), \Bsig).
\end{align*}
Here, the light curve model, $T(\bt,\th)$, is the {\it mean function} of the Gaussian, and the $\Bsig$ is the covariance matrix, where $\Bsig = \mathrm{diag}(\bsig)$. We can maximise this likelihood with respect to the light curve parameters to obtain a point estimate of $\th$, equivalent to minimising the $\chi^2$ statistic if the uncertainties are held fixed. However, using the full likelihood means we can fit for a constant noise term, $\sigma$, or alternatively a scale factor if the flux uncertainties vary with time. This is conceptually similar to setting the reduced $\chi^2$ equal to one, but allows the `uncertainty in the uncertainties' to be taken into account.

To obtain the joint or marginal posterior probability distribution(s) for $\th$, we turn to Bayes' theorem:
\[
p(\th | \D) = \frac{ p(\D | \th) \,p(\th)}{p(\D) },
\]
where $p(\th | \D)$ is the joint posterior probability distribution of $\th$, $p(\D | \th)$ is the likelihood, $p(\th)$ is our prior probability distribution for $\th$, and $p(\D)$ is the evidence under our model, independent of $\th$. In general each of these probabilities should be conditioned on the context, which represents the explicit and implicit assumptions made about our data and models, such as the validity of the transit model and assumption of Gaussian noise, but we usually drop it for clarity. Indeed, for the remainder of this paper Gaussian noise is assumed, which may not always be the case in practice. The consistency of the residuals with a Gaussian distribution can be verified, and generally, rejecting a few outlying points when $N$ is large is a simple and uncontroversial procedure. However, a more detailed treatment of outliers is beyond the scope of this paper.

To infer the value of a single parameter in $\th$, say $\theta_k$ we marginalise over the remaining parameters  $\th_{\backslash k}$:
\[
p(\theta_k | \D) = \int p(\th | \D) d\th_{\backslash k }.
\]
In general this problem is analytically intractable, and we turn to numerical methods such as Markov Chain Monte Carlo (MCMC) to sample from the posterior, where an estimate of the joint and marginal distributions can be easily extracted from the samples, and is routinely used in the exoplanet literature \citep[see \eg][]{Ford_2005,Holman_2006,Cameron_2007,Gibson_2008}.

\subsection{Inference using systematics models}
\label{sect:sys_models}

Unfortunately the light curve is typically degraded by instrumental systematics, and our assumption that the flux values are independent and normally distributed around the light curve model is no longer true. This must be accounted for when inferring $\th$, a problem first discussed in detail by \citet{Pont_2006} in the context of transit surveys. Several methods have been suggested to account for additional systematic noise by rescaling the uncertainties to account for the loss of information \citep{Pont_2006,Winn_2008}, or shuffle the residuals from the best fit model to preserve the correlated noise in the light curves and evaluate the effect on the parameter uncertainties \citep[Residual Permutation, ][]{Gillon_2007}. Alternatively we can use stochastic processes to fit for and account for time-correlated noise such as the wavelet approach of \citet{Carter_2009}, or based on time-dependent Gaussian processes \citep{Gibson_2013a,Gibson_2013b}.

In many cases we instead attempt to model the systematics as a parametric function of a set of auxiliary parameters
 extracted from the data
\citep[often called optical state vectors or decorrelation parameters; see][for a more detailed discussion of linear basis models]{Gibson_2011}. These could be the position of the star on the detector, width of the stellar PSF, temperature of the detector and so on, and the reader is referred to previously mentioned papers for more detail.
This in principle allows us improve the precision of our parameter estimates. Using simple parametric models for systematics has long been a standard approach for studies of exoplanet atmospheres \citep[\eg][]{Brown_2001b,Gilliland_2003,Pont_2007,Swain_2008,Sing_2011,Huitson_2013,Nikolov_2014}.
A common problem with this approach is that there is often no unique and obvious way of modelling the instrumental systematics, and our inference of $\th$ is no longer dependent on simple assumptions like a transit model and Gaussian noise, but rather on the subjective choice of how to model the systematics. When we choose a systematics model, $\S$, for our data, we are effectively conditioning our results on $\S$, and we should rewrite Bayes' theorem accordingly:
\[
p(\th,\ph | \D,\S) = \frac{ p(\D | \th,\ph,\S)\, p(\th,\ph|\S) }{ p(\D|\S) },
\]
where $\bphi$ is an additional parameter vector containing parameters of the systematics model. We can similarly obtain a marginal distribution for $\theta_k$, by also marginalising over the systematics parameters:
\[
p(\theta_k | \D,\S) = \int\int p(\th,\ph | \D, \S)  d\ph\, d\th_{\backslash k }.
\]
Crucially, this takes into account the uncertainty in the parameters of the systematics model, {\it but not the choice of systematics model}.

In this case, our knowledge of $\th$ is explicitly dependent on the choice of $\S$, whether it is an arbitrary choice or objectively selected from a group of models. If $\S$ does not represent the true systematics model, then our inference of $\th$ is unreliable. This is a crucial point to keep in mind for nearly all cases of transmission and emission spectroscopy. Naturally this will also affect any inference of atmospheric composition or structure, which will in turn be conditioned on the systematics model, $\S$.

This is particularly problematic for seemingly arbitrary choices of systematics models. \cite{Gibson_2011} discuss this in the context of HST/NICMOS transmission spectroscopy, where subtly changing the systematics models leads to significantly different transmission spectra and therefore to different physical interpretations of the planetary atmosphere. When an arbitrary choice of systematics model influences the exoplanet spectrum, and therefore the physical interpretation of the data, clearly we must test the assumptions that go into our systematics models in order to gain a more robust result from our data. Otherwise subjective choices will strongly influence our results, and different authors will come to different conclusions from the same data.
More objective inference can be obtained by using model selection criteria to select the most appropriate model as discussed in the following section, and is the approach now followed by the majority of studies using parametric systematics models \citep[\eg][]{Gibson_2010,Sing_2011,Huitson_2013,Nikolov_2014,Crossfield_2013}.

\subsection{Model selection}
\label{sect:model_selection}

Typically, systematics models are constructed from a set of auxiliary data extracted from the observations. The simplest model one can construct is a linear combination of these auxiliary data, and is given by the vector product $\bx_i^T \bphi$, where $\bx_i$ is a vector containing $L$ auxiliary parameters for time $i$, and $\ph$ contains the $L$ weights for each parameter. The likelihood is now given by:
\begin{align*}
 p(\D | \th, \ph, \S ) &= \prod_{i=1}^N \N (f_i | T(t_i,\th) \times \bx_i^T \bphi, \sigma_i^2)\\
 &= \N (\bflux | T(\bt,\th) \circ (\mathbfss{X}\ph), \Bsig),
\end{align*}
where $\mathbfss{X}= (\bx_1,\ldots,\bx_N)^T$ is the $N$ by $L$ design matrix, and $\circ$ denotes element wise multiplication. A vector of ones can be added to $\mathbfss{X}$ to account for an arbitrary offset in the flux, or alternatively absorbed into the light curve model. An additive basis model could also be applied, but here only a multiplicative model is considered.
Often a range of equally plausible basis models could be proposed to explain the systematics. For example we could include only a subset of the auxiliary parameters, or use higher order polynomials of the inputs to represent the systematics.

In the case of multiple systematics models $\S_q$, it is not sufficient to select the model that gives the lowest $\chi^2$, or that which produces the `whitest' residuals.
For mathematically principled model selection based on Bayesian inference, we must instead use the Bayesian evidence, $p(\D|\S)$. This is the denominator of Bayes' theorem in Sect.~\ref{sect:sys_models}. It is the normalisation factor to ensure the posterior integrates to unity, and is interpreted as the probability of obtaining the data given a model $\S$, after integrating over the parameters of the model. It is given by: 
\[
p(\D|\S_q) = \int\int p(\D | \th, \ph, \S_q) p(\th,\ph|\S_q) d\th \, d\ph.
\]

To select our best systematics model, we can calculate a probability for each systematics model and again call on Bayes' theorem:
\[
p(\S_q | \D) = \frac{ p(\D | \S_q) \,p(\S_q)}{p(\D) },
\]
where $p(\S_q)$ is the prior belief of the model. Note that the evidence is often referred to as the marginal likelihood, given it now acts as the likelihood  in Bayes' theorem. In the absence of strong prior belief, the evidence $p(\D | \S_q)$ is the only term that contributes to the probability of $S_q$, and the ratio of these terms (known as the Bayes factor) can be used to assess the relative probabilities of systematics models and select the most appropriate one.

The model evidence naturally selects the model that best explains the data, but with the minimum required complexity; in other words it intrinsically applies Occam's razor to the model selection.
Whilst this term does not come naturally from MCMC methods, one can use alternative numerical methods to evaluate it such as importance sampling \citep[\eg][]{Bishop,Gibson_2013b} or nested sampling \citep{Skilling2004}. One major difficulty is that for the Bayesian evidence to be well defined, we must place proper priors (\ie prior probabilities that integrate to unity) on all the model parameters, or at least those that are not common to all models. In this case any selection of proper priors are also informative priors, and therefore the choice of prior can influence the model selection. An easy to implement approach to this based on Bayesian Linear Basis Models is discussed in Sect.~\ref{sect:laplace}.

A commonly used albeit crude approximation to the evidence is the Bayesian Information Criterion \cite[BIC;][]{Schwarz_1978}. Here the log evidence is approximated by
\begin{align*}
\ln p(\D | \S_q)&\approx-\frac{1}{2} \cdot\mathrm{BIC} = \ln p(\D\,|\,\th_{*},\ph_{*},\S_q) - \frac{M}{2} \ln N,
\end{align*}
where $\th_{*}$ and $\ph_{*}$ are the best fit model parameters, 
$M$ is the number of variable parameters, and N is the number of observations. Alternatively we can replace $-2\ln p(\D\,|\,\th_{*},\ph_{*},\S_q)$ by $\chi^2$ after removing additive constants.
The best fitting model is that with the largest evidence, or lowest BIC.
Note that the additive `constants' depend on the uncertainties ($\bsig$), so BIC model comparison is invalid if calculated directly from the $\chi^2$ and the uncertainties are rescaled for different models; in this case the full likelihood is required.

An alternative information criterion, but less commonly used in the exoplanet literature is the Akaike Information Criterion\footnote{In this paper a modified form of the AIC called $\mathrm{AIC_C}$ is used to correct for finite sample sizes \citep{Hurvich1989}. In this case $\ln p(\D | \S_q) \approx \ln p(\D\,|\,\th_{*},\ph_{*},\S_q) - MN/(N-M-1)$, and is equivalent to the AIC for large $N$.} \cite[AIC;][]{Akaike1974}. It is given by
\begin{align*}
\ln p(\D | \S_q)&\approx-\frac{1}{2} \cdot\mathrm{AIC} = \ln p(\D\,|\,\th_{*},\ph_{*},\S_q) - M
\end{align*}
For large $N$, the AIC imposes a smaller penalty term for extra complexity as compared to the BIC.

\section{Marginalisation over systematics models}
\label{sect:model_marg}

One important caveat with model selection is that it assumes that one of the tested models is in fact the correct model. Given that the sources of instrumental systematics are poorly understood, we should not always assume that the systematics model will be well described by a linear basis model containing the auxiliary inputs and higher order terms. As we are generally only interested in the light curve parameters rather than the specific systematics model, a more appropriate method is to {\it marginalise over the various systematics models} as well as the model parameters. This in principle can take into account the uncertainty associated with the choice of systematics model, provided that a suitable range of models is tested. As will be discussed in Sect.~\ref{sect:gps}, this is one of the motivations for the Gaussian process framework introduced by \citet{Gibson_2012}.

Again using Bayes' theorem, the marginalised distribution of $\btheta$ is given by a sum of the individual (model-specific) distributions of $\th$, weighted on the evidence for each systematics model:
\begin{equation}
\label{eq:marg}
p(\th | \D) = \sum_q p(\th | \S_q, \D )\,p(\S_q | D)
\end{equation}
where
\[
p(\th | \S_q, \D ) = \int p(\D | \th, \ph, \S_q) \,p(\th, \ph | \S_q) \,d\ph,
\]
and is normalised so that $\int p(\th | \S_q, \D ) \,d\th = 1$, 
and $p(\S_q | \D)$ is the marginal evidence as defined in Sect.~\ref{sect:model_selection} (normalised so that $\sum_q p(\S_q | \D) = 1$). If one of the models is heavily preferred over the others, then this result will be equivalent to model selection, but this is important to check during inference.

Even if the underlying distributions of $\th$ conditioned on the individual systematics models were Gaussian, in general the resulting Gaussian mixture model is non-Gaussian. We could work with the full mixture distribution, but in the current work the mean and variance of the mixture distribution will suffice. For a single parameter, these are calculated assuming the conditional distributions are Gaussian with means and standard deviations $m_q$ and $s_q$. In this case the mean $m$ and standard deviation $s$ of the parameter with probability distribution given by the Gaussian mixture distribution are related by:
\[
m = \sum_q p(\S_q | \D)\, m_q,
\]
and
\[
s^2 = {\sum_q p(\S_q | \D) \left[ (m_q - m)^2 + s_q^2 \right]},
\]
and are derived by summing the moments from the individual Gaussian components \citep[\eg][]{Fruhwirth_2006}.

\subsection{Feature space expansion}
\label{sect:feature_space}

Before model selection or marginalisation over multiple models, we first must decide on a set of systematics models to test. Typically, a combination of the basis inputs and higher order terms are explored. For the remainder of this paper, such a {\it feature space} is used to define the set of models $\S_q$. Given a set of measured auxiliary parameters $\mathbfss{X}$, we might wish to test a variety of models using simple polynomial expansions of them, and models that leave out some of the inputs. Assuming that we never use a higher order term without all corresponding lower order terms for a given input, the number of possible models is given by $(1+p)^L$, where $p$ is the highest order of the polynomial expansion and $L$ are the number of auxiliary inputs. For example, if we have two auxiliary inputs $\mathbf{a}$ and $\mathbf{b}$, then a systematics model of third order in $\mathbf{a}$ and second order in $\mathbf{b}$ is given by:
\[
\bx_i^T\ph \triangleq  \phi_1{a_i} + \phi_2{a_i^2} + \phi_3{a_i^3} + \phi_4{b_i}+ \phi_5{b_i^2},
\]
where higher order terms of $\mathbf{a}$ and $\mathbf{b}$ are incorporated into $\bx$ and therefore $\mathbfss{X}$.

For large numbers of auxiliary inputs and/or high expansion orders this can get prohibitively expensive, particularly if using numerical methods to calculate the evidence. Such expansions are commonly used as systematics models \citep[\eg][]{Sing_2011,Huitson_2013,Nikolov_2014}, but are used for model selection rather than marginalisation. We might also consider using cross-terms in the basis expansion, which are used in some systematics models, but this is not considered in the current work.

\subsection{Laplacian approximation of the evidence and shrinkage priors}
\label{sect:laplace}

In Sect.~\ref{sect:model_selection} we discussed the Bayesian evidence and the BIC and AIC as crude approximations to it, often used for model selection. The difficulty in using the full Bayesian evidence is partly that it's relatively slow to calculate using numerical methods, but more importantly that we need to assume proper priors on the parameters that are not common to the models we are comparing. We can improve our evidence approximation by using a Laplacian approximation to the posterior integral \citep[\eg][]{Bishop}.

The Laplacian approximation uses a 2nd order Taylor expansion of the log posterior around the maximum posterior value. The Jacobian is zero at a stationary point so we need only consider the Hessian matrix, which is the negative inverse covariance matrix. We therefore only need an estimate of the maximum (unnormalised) posterior value, and the covariance matrix of our variables $\mathbfss{K}$, both of which are easily obtained from an MCMC or a Levenberg-Marquart optimisation. In this case we are approximating the posterior by a Gaussian distribution, and the integral can be estimated from the ratio of the maximum posterior value to that of an equivalent normalised Gaussian distribution. 
The marginal likelihood in this case is estimated as:
\[
p(\D | \S_q) \approx \frac{ q(\th_{*},\ph_{*} | \D,\S_q) }{ \N(\th_{*},\ph_{*} | \th_{*},\ph_{*},\mathbfss{K})  } = \sqrt{|2\pi \mathbfss{K}|} \cdot q(\th_{*},\ph_{*} | \D,\S_q),
\]
where $q(\th_{*},\ph_{*} | \D,\S_q) = p(\D\,|\,\th_{*},\ph_{*},\S_q)\,p(\th_{*},\ph_{*}|\S_q)$ is the unnormalised posterior. In practice it is better to calculate the log evidence:
\begin{align*}
\ln p(\D | \S_q) &\approx \ln q(\th_{*},\ph_{*} | \D,\S_q) + \frac{M}{2} \log{2\pi} + \frac{1}{2}\ln |\mathbfss{K}| \\
&\approx \ln q(\th_{*},\ph_{*} | \D,\S_q) + \frac{1}{2}\ln |2\pi\mathbfss{K}|
\end{align*}

This is in many ways resembles the BIC in that increasing the number of data points or the model complexity typically increases a penalty term  \citep[indeed the BIC and AIC can be derived following these arguments using a Gaussian approximation and vague priors, \eg][]{Bishop}, but uses the covariance matrix to obtain a more accurate estimate of the evidence. For the models discussed in this paper, the Laplace approximation provides a very good estimate of the log evidence, as verified from importance sampling using the technique described in \citet{Gibson_2013b}.

In order to apply priors to the non-common systematics parameters, we follow a Bayesian treatment of linear regression, and introduce zero mean priors to each of the basis model weights \citep{Bishop}. In this case we need only assign proper priors to the non-common inputs, and this is where the Laplace approximation shows its value. Instead of using simple model selection (or marginalisation) to choose the systematics models, we can use more complex models providing that we have shrinkage priors assigned to each additional basis vector.
This is the general principle behind Bayesian Linear Basis Models. This allows linear models with large numbers of basis functions to be fit to relatively few data points, whilst mitigating over-fitting. This is because the shrinkage priors restrict the complexity of the model, by avoiding the model being able to fine-tune itself using large basis weights. The width of the priors cannot be set by optimising the likelihood or posterior, but can be set through optimising the marginal likelihood. The reader is referred to \citep{Bishop} for more details.

To implement this for our systematics models, for each of the systematics parameters ($\ph$), we can assign a zero mean Gaussian prior with variance $\alpha^2$:
\[
p(\ph | \S_q) = \N(\bz,\alpha^{2} \mathbfss{I})
\]
We first use an MCMC to determine the mean and covariance for the likelihood function (\ie with posterior uniform priors), and approximate it as a Gaussian. We can then {\it analytically} multiply the likelihood and prior as they are both Gaussian\footnote{For convenience a very broad Gaussian distribution for each of the variable light curve parameters is taken as the prior. These do not affect the mean and variance of the posterior, and is equivalent to adding an arbitrary constant to the log evidence of each model.}, and use the Laplace approximation to evaluate the log evidence. Thus by using the Laplace approximation we can easily maximise the evidence with respect to $\alpha$ to find suitable shrinkage priors for our inference. We can subsequently use the posterior distribution with the maximised value for $\alpha$ to obtain our desired distribution for $\th$, perform another MCMC analysis using priors on the systematics parameters fixed to $\alpha$, or again use the values for the log evidence for each tested model to find the best fit model. Alternatively, we can marginalise over the feature space to obtain a marginalised distribution for $\th$.

\subsection{Gaussian processes as marginalisation over systematics models}
\label{sect:gps}

A Gaussian process (GP) can be seen as a convenient shortcut to the above marginalisation over many systematics models.
Formally, a GP is an infinite collection of random variables, any finite number of which have a joint Gaussian distribution \citep{Rasmussen_Williams,Bishop}.
They were introduced for modelling time-series light curves by \citet{Gibson_2012} and have been subsequently used in various forms in \citet{Gibson_2012b,Aigrain_2012,Gibson_2013a,Gibson_2013b,Evans_2013}, which the reader is referred to for further details. 

A GP may be derived from a probabilistic linear basis model, similar to those discussed in the previous section, by considering the mean and covariance of a random draw from the model. To make the connection between GPs and feature space models discussed earlier, we follow the derivation of \citet{Rasmussen_Williams} and \citet{Bishop}, starting with a basis model as described in Sect.~\ref{sect:model_selection}:
\[
\bflux = \mathbfss{X}\bphi + \bep,
\]
where $\bep$ is the white noise term distributed according to $\bsig$, i.e. $p(\bep) = \N(\0,\sigma^2\I)$, and $\sigma^2$ is the variance of the white noise. To generate a probability distribution over the resulting functions, we must place a prior distribution over the basis weights, $\bphi$. Similarly to the Laplace approximation, we place a zero mean Gaussian with prior variance $\alpha^2$:
\[
p(\ph) = \N(\bz,\alpha^{2} \I).
\]
Following the derivation of \citet{Rasmussen_Williams,Bishop}, we then proceed to find the joint probability distribution of $\bflux$, which is also Gaussian, and therefore we only need to calculate the mean, $\bmu$ and covariance, $\Bsig$. The mean function is given by the expectation of $\bflux$:
\[
\bmu = \E [\bflux] = \E [ \mathbfss{X}\bphi] + \E[\bep] = \mathbf{0},
\]
and the covariance matrix is given by the expectation of $(\bflux-\bmu)(\bflux-\bmu)^T$:
\begin{align*}
\Bsig &= \E [(\bflux-\bmu)(\bflux-\bmu)^T] = \E [(\mathbfss{X}\bphi + \bep)(\mathbfss{X}\bphi + \bep)^T]\\
&=\alpha^2 \mathbfss{X}\mathbfss{X}^T + \sigma^2 I
\end{align*}
Our GP is fully defined by its mean and covariance, and is now given by:
\[
p(\bflux) = \N(\0,\alpha^2 \mathbfss{X}\mathbfss{X}^T + \sigma_N^2 I).
\]

Using this procedure, we can write any Bayesian linear basis model as a GP. Similarly, we can directly define the kernel function of a GP, and convert it back into feature space. This is where GPs become extremely powerful for non-parametric data analysis, as we can define kernels that have infinite dimensions in feature space.  A Gaussian process can therefore be used to place a probability distribution over a huge class of functions, and therefore automatically marginalise over many systematics models simultaneously. This was the motivation behind the GP systematics framework introduced by \citet{Gibson_2012}. For example, the squared exponential kernel used by \citet{Gibson_2012} can be represented by an infinite number of basis functions \citep{Rasmussen_Williams}.

To apply GPs to exoplanet light curves, a mean function representing the light curve model can be easily included in the GP, and the kernel function (the stochastic component) represents the systematics model. In this case the likelihood function for a GP is identical to that given in Sect.~\ref{sect:inference}:
\begin{align*}
 p(\D | \th, \bphi) &=\N (\bflux | T(\bt,\th), \Bsig).
\end{align*}
The difference is that the covariance matrix is no longer a diagonal matrix, but is given by the kernel function which defines each element $\Bsig_{ij}$ in $\Bsig$. The squared exponential kernel used in \citet{Gibson_2012} is given by:
\[
\mathbf\Sigma_{ij} = k(\bmath{x}_i, \bmath{x}_j) = \xi^2 \exp\left[-\sum_{k=1}^K\eta_k(x_{i,k} - x_{j,k})^2\right] + \delta_{ij} \sigma^2.
\]
Here $\bmath{\eta} = (\eta_{1} ,\ldots, \eta_{K})^T$ are the inverse length scale parameters, one for each input in $\mathbfss{X}$, $\xi$ is the height scale, and $\delta_{ij}$ is the Kronecker delta. Note that the parameters of our mean function and covariance matrix are often referred to as {\it hyperparameters} in the context of Gaussian processes. Two inputs are highly correlated when {\it all} inputs in $\bx$ are similar (with similarity governed by $\bmath\eta$), and uncorrelated otherwise. This therefore represents a joint systematics model, \ie the systematics are not independent functions of the components of $\mathbfss{X}$ added together.

We can make the systematics components independent by modifying the kernel so that each input has its own, independent height scale:
\[
\mathbf\Sigma_{ij} = k(\bmath{x}_i, \bmath{x}_j) = \sum_{k=1}^K \xi_i^2 \exp\left[-\,\eta_k(x_{i,k} - x_{j,k})^2\right] + \delta_{ij} \sigma^2,
\]
This kernel now represents a systematics model made up of independent functions of the inputs in \mathbfss{X}. The first kernel is hereafter referred to as the squared exponential (SE) kernel, and the second as the summed squared exponential (SSE) kernel.

Inference of our light curve parameters proceeds identically to before, where we numerically sample from the joint posterior probability distribution, and parameters in the covariance kernel are treated in just the same way as our light curve parameters, or the systematics parameters of the deterministic models. For this work, a logarithmic prior is imposed on the length scales $\bmath{\eta}$ of the form:
\[
p(x) = \left \{
\begin{array}{ll}
1/x & \mathrm{if}~x>0\\
0 & \mathrm{if}~x\leq0
\end{array}
\right.
\]
These are improper priors, and are equivalent to re-parameterising $\bmath{\eta}$ as $\log \bmath{\eta}$, and is the natural parameterisation for such a scale parameter. This also acts as a shrinkage prior, as it encourages the inverse length scale to be small unless required to explain the data.
However, even without priors of this type, GPs are intrinsically Bayesian and try to find the simplest explanation for the data, as a complexity penalty is introduced in the first $\Bsig$ in the Gaussian likelihood.

GPs are an extremely effective and flexible method for modelling systematics; however, this comes at a cost. For each evaluation of the likelihood function we must invert the covariance matrix\footnote{In practice the matrix equation $\mathbfss{A}\bx = \mathbf{b}$ is solved using Cholesky decomposition, where $\mathbfss{A}$ is a positive semi-definite matrix.}. For regular inputs \citep[such as the time-dependent GPs considered in][]{Gibson_2013a,Gibson_2013b}, this can be sped up considerablely when the covariance matrix is Toeplitz \citep[\eg][]{Trench_1964}. In general for systematics models of $\mathbfss{X}$ we must invert the full covariance matrix.

\section{Simulations of transit light curves with instrumental systematics}
\label{sect:simulations}

In order to investigate the best practices for inferring light curve parameters in the presence of systematics, a series of transit light curves were simulated and degraded with simulated systematics, and their parameters were recovered using a variety of systematics modelling techniques, based on those discussed in Sect.~\ref{sect:inference}. The transit models were simulated assuming a circular orbit, with parameters given in Tab.~\ref{tab:transit_parameters}. They were generated using the analytic models of \citet{Mandel_Agol_2002}, using a quadratic limb darkening law with parameters $c_1$ and $c_2$. The simulated light curves were of typical hot Jupiter-like transits, with the transit and noise parameters for each model picked from a uniform distribution from the (arbitrary) ranges outlined in Tab.~\ref{tab:transit_parameters}.

\begin{figure*}
\centering
\includegraphics[width=175mm]{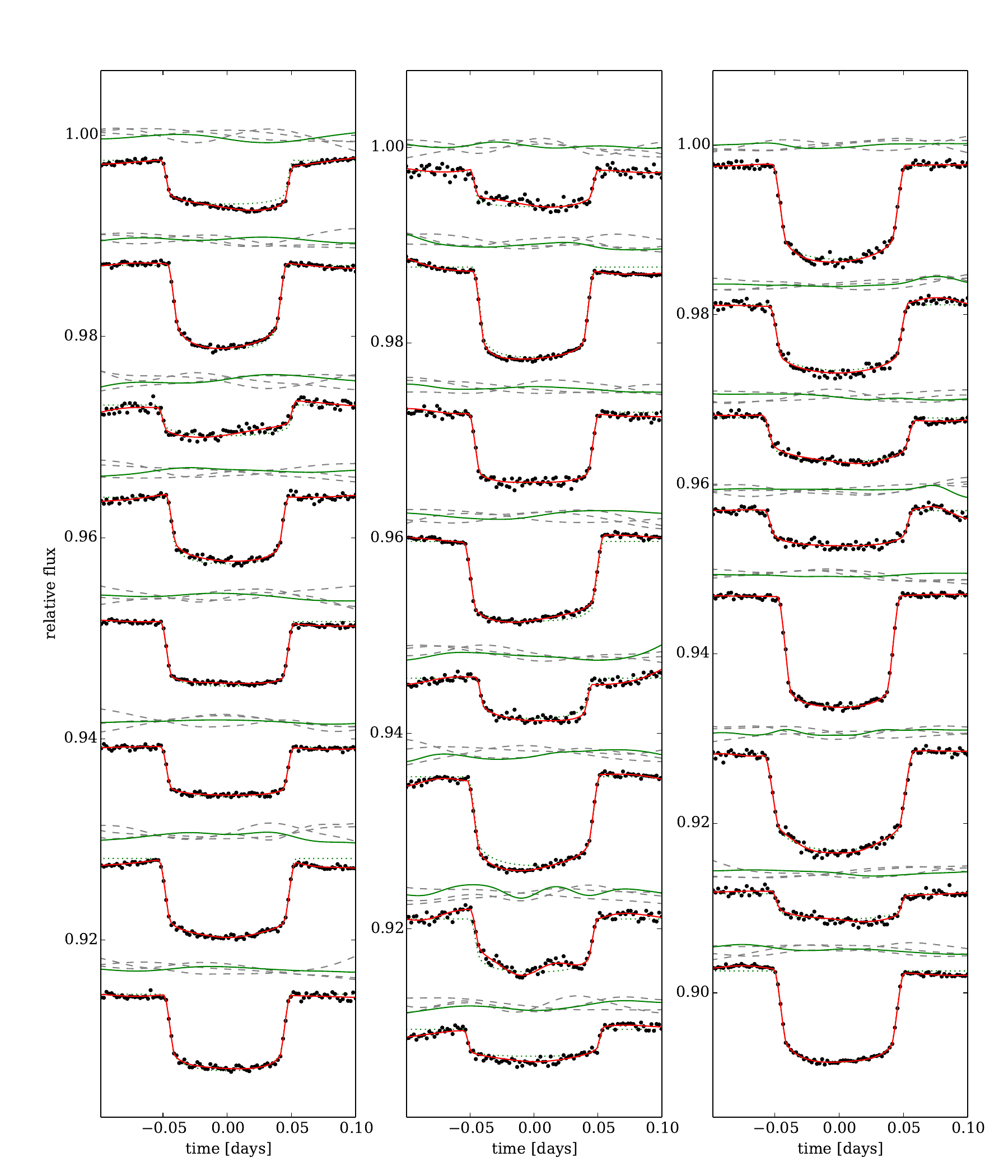}
\caption{Examples of simulated light curves generated to test the fitting techniques. The red line is the model light curve, the dashed lines are basis inputs used to generate the systematics, and the green lines are the resulting systematics function. The left, centre and right panels are light curves generated with the linear systematics model (Sect.~\ref{sect:lbm}), the feature space systematics model (Sect.~\ref{sect:fs}), and the stochastic systematics model (Sect.~\ref{sect:gp_fits}), respectively. For each set of simulations, 10\,000 light curves were generated and fitted to recover the planet-to-star radius ratio.}
\label{fig:lcvs}
\end{figure*}

Next, a set of stochastic basis models \mathbfss{X} were generated by taking draws from a Gaussian process with a rational quadratic kernel \citep{Rasmussen_Williams}:
\[
k(t_{i},t_j) = \exp\left(1+\frac{(t_i-t_j)^2}{2\beta l^2}\right) ^{-\beta},
\]
where $t_i$ and $t_j$ are the times of the two inputs, $l$ is the (temporal) length scale and $\beta \in (0,\infty)$.
This can be interpreted as a sum of squared exponential kernels of varying length scales, and becomes the squared exponential kernel when $\beta \to \infty$.
The kernel parameters were fixed at $l=0.07$ and $\beta=0.1$; these were chosen to generate `realistic' looking, smooth input bases with typically $2-4$ turning points over the transit duration. Examples of generated basis functions are shown in Fig.~\ref{fig:lcvs}.

Each basis was first scaled to have mean 0 and variance 1. In some cases discussed below, some inputs from the generated basis set were excluded from the instrument model, or higher order terms were included to generate a feature space (\ie the basis inputs were raised to integer powers and added to the design matrix $\mathbfss{X}$).
For each basis set, the weight parameters ($\ph$) were randomly generated from a uniform distribution with range (-1,1), and the systematics model was generated as $\mathbfss{X}\ph$. The resulting systematic function was scaled and shifted to have mean 1 and standard deviation (or `red noise') $\sigma_r$. The systematic function and transit function were then multiplied together, and white noise was added with standard deviation $\sigma_w$ to generate the final simulated light curve. 

For each simulated light curve, the planet-to-star radius ratio $\rho$ was recovered using a variety of systematics models outlined below, with all other transit parameters fixed except for $f_{oot}$, the out-of-transit flux, which was allowed to vary freely. All parameters of the systematics model were allowed to vary, as was the white noise. For each systematics model, the fitted planet-to-star radius ratio ($\hat\rho$), its uncertainty ($\sigma_\rho$), and also the true value ($\rho$) were recorded for each transit fit. $\hat\rho$ and $\sigma_\rho$ were found using an MCMC \citep[see][and references therein for more details of the implementation used here]{Gibson_2012}. For each model fit, two MCMC chains were run in parallel. After an initial burn in period (typically of length 5000, but increased for more complex models), the chains were continued until the Gelman \& Rubin statistic \citep{GelmanRubin_1992} was below 1.01 for all variable parameters, or a maximum chain length was reached. In almost all cases the chains converged, but this was not confirmed for every single run.

To compare results, the reduced $\chi^2$ ($\chi^2_\mathrm{r}$) for each systematics modelling technique was calculated, which should be distributed according to a normal distribution with mean one and standard deviation $\sqrt{2/N_t}$ for large samples, where $N_t$ are the number of transit light curves fitted. The `number of sigma' statistic \citep{Carter_2009} was also calculated:
$
\N_\mathrm{s} = (\hat\rho-\rho)/\sigma_\rho.
$
This should be distributed with a mean of zero and variance of unity if the distribution of $\hat\rho$ is normally distributed around its true value with the derived uncertainty. These are both measures of the reliability of the inferred probability distribution of $\rho$. Models were also compared using the mean uncertainty $\bar\sigma_\rho$ and the mean accuracy (\ie mean of $\left|\hat\rho-\rho\right|$). The favoured systematics model is the one with the lowest mean uncertainty and accuracy but also with reliable uncertainties. For the remainder of this paper, an acceptable fit is roughly defined as one with a mean uncertainty within $\approx$10\% of the true systematics model, and reliable uncertainties when $\chi^2_r$ is less than $\approx$1.1.

Unless stated otherwise, 100 data points were generated over a 4.8 hour duration. This is at the low end of the number of data points in a typical light curve, but was chosen to speed up inference over thousands of light curves. However, the resulting cadence of $\approx3$\,mins is similar to typical \HST/STIS light curves, and in principle systematics modelling should become easier with higher cadence. The results of these simulations are only strictly valid for the range of transit and systematics parameters chosen here, but nonetheless such simulations should be able to establish some sensible guidelines for inferring light curve parameters in the presence of instrumental systematics. Examples of the resulting basis functions and light curves are shown in Fig.~\ref{fig:lcvs} for a variety of the light curve simulation methods outlined below.

\begin{table}
\caption{Transit and basis function parameter distributions used to produce simulated light curves.}
\label{tab:transit_parameters}
\begin{tabular}{lll}
\hline
\noalign{\smallskip}
Parameter & Symbol & Range\\
\hline
\noalign{\smallskip}
Central transit time (days) & $T_0$      &  0 \\
Period (days) & $P$    &   (2.75,3.25)\\
Scaled semi-major axis & $a/R_\star$    &   (9,11)\\
Impact parameter & $b$    &   (0.05,0.25)\\
Planet-to-star radius ratio & $\rho=R_p/R_\star$    &   (0.08,0.12)\\
Limb darkening parameters & $c1,c2$    &   (0.1,0.3)\\
White noise & $\sigma_w$    &   (0.0001,0.0004)\\
Red noise & $\sigma_r$    &   (0.0001,0.0004)\\
\hline
\noalign{\smallskip}
\end{tabular}
\end{table}

\subsection{Linear systematics model}
\label{sect:lbm}

The first set of simulated light curves were generated using a linear combination of known basis functions for the systematics. For each simulated transit, five basis functions were generated, but only the first three were used to create the systematics models as described in the previous section. This is a realistic scenario where many auxiliary inputs are extracted, but it is unclear which ones are the primary cause of the systematics (although the assumption of linearity is typically unrealistic). Ten thousand simulated light curves were generated in total, and the transit depth was recovered using a series of systematics models.

These included simple linear basis models (LBM), using the correct, known basis expansion (LBM: correct exp), using only two of the correct bases (LBM: 2 inputs), and finally all five of the bases (LBM: 5 inputs). A series of 1st order feature space (FS) expansions of the first four linear bases in $\mathbfss{X}$ were also used to fit each light curve, in this case resulting in 16 combinations of basis models.
The distributions for the light curves parameters were then obtained by marginalising over the various feature space models using Eq.~\ref{eq:marg}. The evidence was estimated using the BIC (FS: marginalised BIC), the \AICC (FS: marginalised AIC), and the Laplace approximation to the evidence combined with shrinkage priors (FS: marginalised LP) as described in Sect.~\ref{sect:laplace}. The statistics were also recorded for the best fit systematics models found using each evidence approximation. Finally, a Laplacian approximation was used to compute a model with all five input bases, and after the value of $\alpha$ was determined by maximising the evidence, it was fixed and the posterior was resampled using MCMC (LP: 5 inputs).

The results of using the various systematics models are shown in Table~\ref{tab:lbm_results}, including $\chi_\mathrm{r}^2$, the mean of the $\N_\mathrm{s}$ distribution, the mean uncertainty, the mean accuracy, and the rescaling of uncertainties required to set the $\chi_\mathrm{r}^2$ equal to 1. Note the standard deviation of $\N_\mathrm{s}$ is given by $\sqrt{\chi^2_r}$, if the mean of $\N_\mathrm{s}$ is exactly zero. Clearly the best model to use for recovery of the planet-to-star radius ratio is the correct, known basis model, which generates ideal statistics and results in reliable uncertainties. However, in general the systematics model is not known {\it a priori}. Using more complex systematics models which contain the correct model as a subset (\eg LBM: 5 inputs) also results in reliable uncertainties, but at the cost of inflated uncertainties. Therefore in order to get the most from our data, model selection or marginalisation is required.

\begin{figure*}
\centering
\includegraphics[width=170mm]{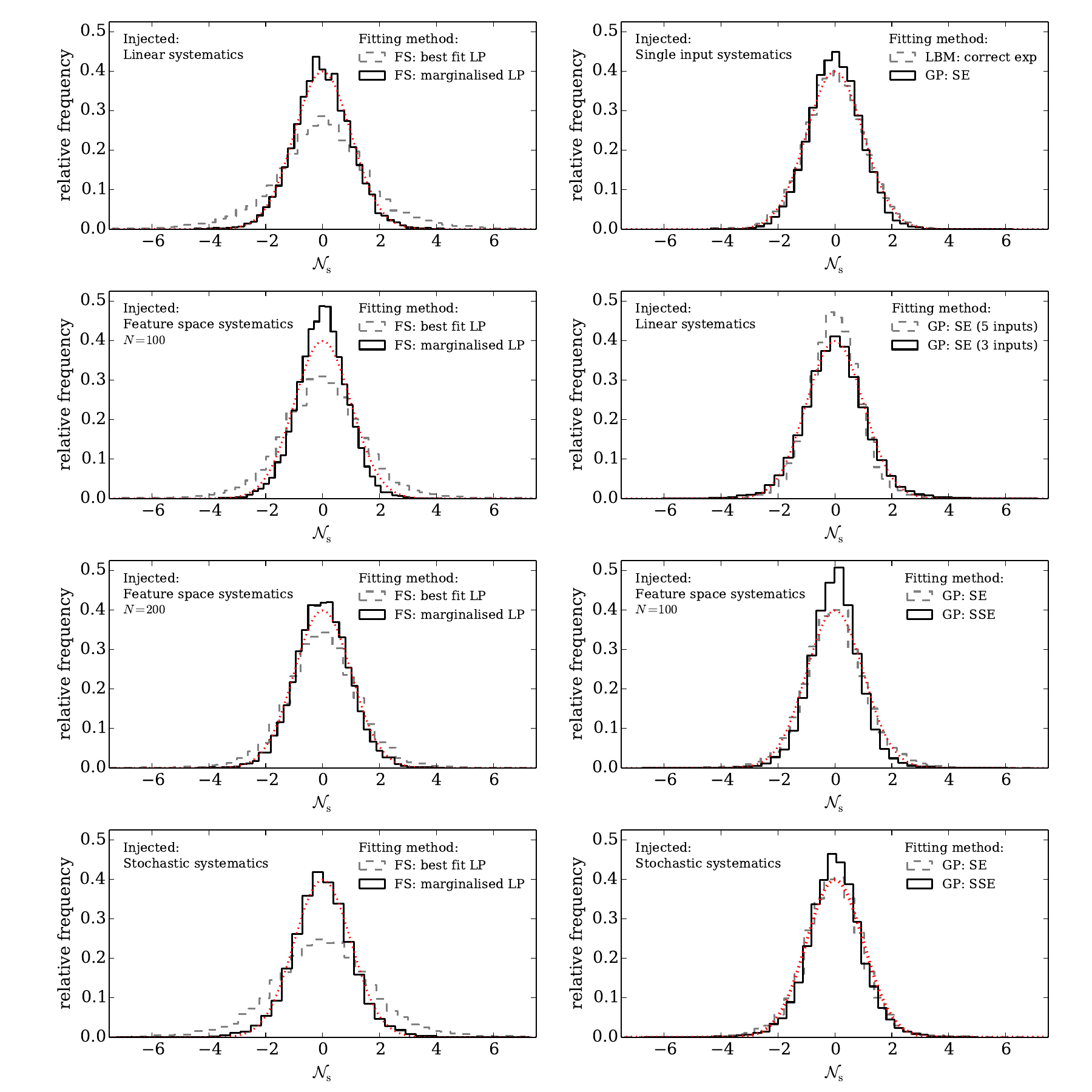}
\caption{Histograms of the number-of-sigma ($\N_\mathrm{s}$) statistic for a selection of the simulated light curves. The systematics model and the fitting methods are marked on the plots. The left panels show the difference between using the best-fit feature space expansion and marginalising over the full feature space, showing that in all cases marginalisation provides a more reliable recovery of the planet-to-star radius ratio. The right panel shows the results for the Gaussian process fits to various systematics models, showing that in all cases the GP models provide a good model for the systematics.}
\label{fig:histograms}
\end{figure*}

From the feature space models, it is clear that marginalising over the feature space models is better than selecting one in all cases. This is because marginalisation will help reduce the effects of poor model selection, and in particular is important when the evidence from multiple feature space models does not clearly select one over the others. Fig.~\ref{fig:histograms} shows $\N_\mathrm{s}$ distributions for the best fit and marginalisation over feature space using the Laplace approximation.
It is clear that marginalisation over feature space gives $\N_\mathrm{s}$ distributions that are closer to a unit normal distribution than model selection. It is also clear that the \AICC\ is better than the BIC. This is due to the \AICC\ preferring more complex models, which results in more reliable yet larger uncertainties. This is a desirable quality for selecting or marginalising over systematics models, as we are not particularly interested in which model is most likely to accurately describe the systematics models, but prefer to pick the least complex one that gives us reliable uncertainties. The Laplace approximation to the evidence provides the best model selection and marginalisation criterion, but is only slightly better than the AIC here, and with significantly more computation required. Finally, the Laplace approximation using shrinkage priors and a more complex model gives reasonable uncertainties along with a modest increase in the uncertainties. This is a good balance between allowing model complexity and preventing over-fitting, but it is still dependent on the specific basis inputs used, so methods using model selection or marginalisation are generally preferred.

Perhaps the most noteworthy result is that no systematics model selection procedure results in reliable statistics, {\it even when we know that the correct systematics model is within the tested models}. Most strikingly, the commonly used method of BIC model selection (at least in this case) typically underestimates the uncertainties by a factor of $\sim$100\%, although this is likely not always the case in practice (Sect.~\ref{sect:practical}).

\begin{table*}
\caption{Results from fitting 10\,000 light curves simulated using the linear systematics model described in Sect.~\ref{sect:lbm}, \ie a linear combination of three basis functions, and fitting using a combination of those three basis inputs and a further two inputs. The light curves were fitted using linear basis models (LBM), feature space (FS) expansions, and Laplacian approximations using shrinkage priors on the basis weights (LP). The feature space evidence was calculated using the BIC, AIC, and a Laplacian expansion with shrinkage. $\chi_r^2$, the mean of $\N_\mathrm{s}$ ($\bar\N_\mathrm{s}$), the mean uncertainty ($\bar\sigma_\rho$), mean accuracy (mean acc) and the rescaling of the uncertainties required to set $\chi_r^2=1$ are reported for the various fitting methods.}
\label{tab:lbm_results}

\begin{tabular}{lccccc}
\hline
\noalign{\smallskip}
Fitting method & $\chi^2_{\rm r}$ & $\bar{\N_\mathrm{s}}$ & $\bar\sigma_\rho$ & mean acc & rescaling (\%)\\
\hline
\noalign{\smallskip}
LBM: 2 inputs        &   8.76  &   -0.004 &    0.00044 &    0.00080 &      196.0\\
LBM: correct exp  &   1.05  &   -0.016 &    0.00054 &    0.00044 &        2.5\\
LBM: 5 inputs        &   1.02  &   -0.022 &    0.00082 &    0.00065 &        1.2\\
FS: best-fit BIC     &   4.73  &   -0.022 &    0.00036 &    0.00054 &      117.6\\
FS: best-fit AIC     &   3.38  &   -0.020 &    0.00041 &    0.00052 &       83.9\\
FS: best-fit LP      &   3.29  &   -0.026 &    0.00039 &    0.00048 &       81.3\\
FS: marginalised BIC &   1.40  &   -0.014 &    0.00055 &    0.00048 &       18.3\\
FS: marginalised AIC &   1.06  &   -0.014 &    0.00059 &    0.00047 &        3.1\\
FS: marginalised LP  &   1.00  &   -0.015 &    0.00060 &    0.00046 &       -0.1\\
LP: 5 inputs   &   1.38  &   -0.016 &    0.00056 &    0.00050 &       17.5\\
\hline
\noalign{\smallskip}
\end{tabular}

\end{table*}

\subsection{Feature space systematics}
\label{sect:fs}

An additional two sets of simulated light curves were generated using a feature space expansion of basis inputs. Two linear basis functions were generated as before, and higher order terms were added to the design matrix $\mathbfss{X}$. The order for expansion was selected for each basis at random from the range 1--3 for each simulated light curve. The first set of simulations used $N = 100$ data points as in the previous section, and the second set was generated using $N=200$ data points to test the systematics models with higher cadence.

For both sets of simulations, ten thousand light curves were generated, and $\rho$ was recovered using a similar set of techniques as before. 
These methods included using only a linear basis expansion (LBM: linear exp), plus the known, correct basis expansion for each light curve (LBM: correct exp).
A feature space expansion for up to fourth order in both bases was used, resulting in 25 combinations of basis models.
The BIC (FS: marginalised BIC), \AICC\ (FS: marginalised AIC) and Laplace approximations to the evidence (FS: marginalised LP) were then used as before to recover the fitted parameters, as well as the best fit model in each case.
Laplace approximations were also recorded for up to second order and third order expansions (LP: 2nd/3rd order exp), including refits after fixing the priors. Results for the same expansions were also recorded without the shrinkage priors on the weight parameters (LBM: 2nd/3rd order exp). The results for both sets of simulations are shown in Tables~\ref{tab:fs1_results} and \ref{tab:fs2_results}.

In both cases we find similar results to the linear systematics simulations. The best model is unsurprisingly using the correct, known basis expansion. Marginalisation over many models is again preferable to model selection, and the $\N_\mathrm{s}$ distributions for the Laplacian approximation are shown in Fig.~\ref{fig:histograms}, which again show that the marginalisation results in a $\N_\mathrm{s}$ distributions closer to the unit normal. Similarly, using the BIC for model selection typically underestimates the uncertainty on average by $\sim100\%$.
The \AICC\ is perhaps the best model selection criterion here, as the Laplace approximation slightly overestimates the uncertainties, although only by a small factor. Both methods provide reliable recovery of the planet-to-star radius ratio, although the \AICC\ is significantly easier to calculate. Qualitatively, the simulations for $N=200$ produce the same results as for $N=100$ (\ie the same model selection/marginalisation techniques produce the best results), but quantitatively the model selection techniques are more reliable than before. This is not surprising, as model selection should become easier as the number of data points and data quality increases.

\begin{table*}
\caption{Results from 10\,000 simulated using the first set of simulations ($N=100$) from the feature space systematics model, where two basis inputs were used. The basis expansion for each input was selected randomly between first and third order. These were fitted using similar models to those in Tab.~\ref{tab:lbm_results}.}
\label{tab:fs1_results}

\begin{tabular}{lccccc}
\noalign{\smallskip}
\hline
Fitting method & $\chi^2_{\rm r}$ & $\bar{\N_\mathrm{s}}$ & $\bar\sigma_\rho$ & mean acc &  rescaling (\%)\\
\hline
\noalign{\smallskip}
LBM: linear exp      &   5.99  &    0.036 &    0.00046 &    0.00076 &      144.7\\
LBM: correct exp     &   1.04 &   -0.004 &    0.00058 &    0.00046 &        2.2\\
LBM: 2nd order exp        &   1.74  &   -0.008 &    0.00060 &    0.00058 &       31.9\\
LBM: 3rd order exp        &   1.05 &   -0.007 &    0.00076 &    0.00062 &        2.7\\
FS: best fit BIC           &   3.64  &    0.008 &    0.00039 &    0.00053 &       90.7\\
FS: best fit AIC           &   2.65  &   -0.006 &    0.00045 &    0.00055 &       62.7\\
FS: best fit LP            &   2.31  &   -0.008 &    0.00040 &    0.00044 &       51.9\\
FS: marginalised BIC       &   1.36  &   -0.002 &    0.00054 &    0.00048 &       16.5\\
FS: marginalised AIC       &   0.94  &   -0.009 &    0.00065 &    0.00049 &       -3.1\\
FS: marginalised LP        &   0.77  &   -0.011 &    0.00071 &    0.00046 &      -12.1\\
LP: 2nd order exp  &   1.93 &    0.001 &    0.00048 &    0.00049 &       38.9\\
LP: 3rd order exp  &   1.55  &    0.002 &    0.00046 &    0.00044 &       24.3\\
\hline
\noalign{\smallskip}
\end{tabular}

\end{table*}

\begin{table*}
\caption{Same as Tab.~\ref{tab:fs1_results} for the second set of simulations using feature space systematics with $N=200$.}
\label{tab:fs2_results}

\begin{tabular}{lccccc}
\hline
\noalign{\smallskip}
Fitting method & $\chi^2_{\rm r}$ & $\bar{\N_\mathrm{s}}$ & $\bar\sigma_\rho$ & mean acc & rescaling (\%)\\
\hline
\noalign{\smallskip}
LBM: linear exp     &   8.08 &    0.006 &    0.00030 &    0.00058 &      184.2\\
LBM: correct exp    &   1.01 &    0.007 &    0.00033 &    0.00026 &        0.6\\
LBM: 2nd order exp       &   2.17 &   -0.016 &    0.00034 &    0.00037 &       47.2\\
LBM: 3rd order exp       &   1.01 &    0.001 &    0.00040 &    0.00032 &        0.7\\
FS: best fit BIC          &   2.67 &   -0.007 &    0.00027 &    0.00031 &       63.3\\
FS: best fit AIC          &   1.80 &    0.005 &    0.00030 &    0.00031 &       34.0\\
FS: best fit LP           &   1.79 &   -0.006 &    0.00027 &    0.00027 &       33.7\\
FS: marginalised BIC      &   1.36  &   -0.004 &    0.00033 &    0.00029 &       16.5\\
FS: marginalised AIC      &   0.95 &   -0.003 &    0.00037 &    0.00029 &       -2.5\\
FS: marginalised LP       &   0.91  &   -0.001 &    0.00038 &    0.00027 &       -4.6\\
LP: 2nd order exp &   2.21 &   -0.014 &    0.00031 &    0.00034 &       48.8\\
LP: 3rd order exp  &   1.37 &   -0.008 &    0.00031 &    0.00028 &       17.0\\
\hline
\noalign{\smallskip}
\end{tabular}

\end{table*}

\subsection{Gaussian process fitting}
\label{sect:gp_fits}

In order to establish the reliability of GPs for recovering light curve parameters when the systematics can be well described by simple basis models, a GP model was used to recover the planet-to-star radius ratio of similar systematics simulations. These fits were performed separately from the linear basis and feature space fits described earlier, but used the same sets of parameters to generate the light curves. In order to recover the planet-to-star radius ratio, a squared exponential kernel was used as described in Sect.~\ref{sect:gps}, using both the SE and SSE forms as outlined below. A uniform prior was used for the height scale, and a logarithmic prior was used for the inverse length scales, in other words we fitted for $\log \bmath\eta$ instead of $\bmath\eta$ as discussed in Sect.~\ref{sect:gps}. A logarithmic prior was also tried for the height scale, but this was found to be too aggressive. For the majority of cases this didn't affect the fits; however, it was found to produce more outliers in the distribution for $N_\mathrm{s}$, presumably when the prior suppressed modelling of the systematics. Strictly speaking, a logarithmic prior should also be imposed on the height scale, but it proved difficult to check the details of individual fits over thousands of light curves, and we proceeded using uniform priors on $\xi$. As the posterior distribution is no longer well approximated by a Gaussian distribution, an affine-invariant MCMC was used instead of the Metropolis-Hastings algorithm \citep{Goodman_Weare_2010, Foreman-Mackey_2013}, with two chains computed for each fit and the same convergence tests were applied as before.

First a set of ten thousand simulated light curves were generated using only one input basis in $\mathbfss{X}$, with the order of expansion selected at random from the range 1--3, hereafter referred to as the single input systematics model. The fits were performed using a GP with SE kernel (GP: SE kernel) as well as for the correct basis expansion (LBM: correct exp) to recover $\hat\rho$ and $\sigma_\rho$. The second set of simulated light curves were identical to the linear basis systematics described in Sect.~\ref{sect:lbm}. Here, a GP with SE kernels were used with both 3 and 5 of the basis inputs (GP: SE kernel 3/5 inputs). The final set of simulated light curves were identical to that described in Sect.~\ref{sect:fs} (with $N=100$), using 2 basis inputs and an expansion for each selected at random from 1--3. These were fitted using a GP with both SE and SSE kernels (GP: SE/SSE kernel). The results are shown in Tab.~\ref{tab:gp_results}, and a selection of $\N_\mathrm{s}$ distributions are shown in Fig.~\ref{fig:histograms}.

In general the GP models all produce acceptable fits to the light curves, and provide reliable uncertainties. This is despite the GP models not always being ideal models for the data. First, the GP used here is additive, and the true systematics models are multiplicative. Secondly, the SE kernel is not the correct choice for the models presented here, as we know {\it a priori} that the systematics models for each input basis are independent, whereas the SE kernel assumes the systematics are a joint function of the inputs, \ie that all (relevant) inputs must be similar for the covariance to be high. This results in larger uncertainties than the corresponding marginalisation over feature space, as is the case for the linear systematics models. The SSE kernel is the more appropriate choice for independent systematics, but it was not tried for the linear systematics model as the number of hyperparameters increased significantly and made inference over thousands of light curves infeasible. For the feature space systematics, the SSE kernel performs better, and provides almost identical results to the feature space marginalisation with Laplacian approximation. This is despite it being far more flexible than even marginalisation over feature space modes, and required only one posterior distribution to be computed. Forcing the systematics model to be a simple deterministic function of the basis inputs is perhaps a better method for fitting systematics when the true systematics model is contained in the tested model space. In general however, this is quite a rigid assumption, as the true systematics model is very unlikely to be precisely described by a simple polynomial expansion of the inputs, suggesting that GPs are the better choice as they provide greater flexibility and are easier to implement. The following section tests the use of deterministic models on a more general class of systematics models.

\begin{table*}
\caption{Results for 10\,000 light curves simulated using the range of models described in Sect.~\ref{sect:gp_fits}, and recovered using various GP systematics models.}
\label{tab:gp_results}

\begin{tabular}{lccccc}
\hline
\noalign{\smallskip}
Fitting method & $\chi^2_{\rm r}$ & $\bar{\N_\mathrm{s}}$ & $\bar\sigma_\rho$ & mean acc &  rescaling (\%)\\
\hline
\noalign{\smallskip}
Single input systematics model\\
~LBM: correct exp   &   1.02  &    0.002 &    0.00037 &    0.00029 &        1.1\\
~GP: SE kernel   &   0.84 &    0.004 &    0.00045 &    0.00033 &       -8.4\\
Linear systematics model\\
~GP: SE kernel (3 inputs)   &   1.21 &   -0.000 &    0.00070 &    0.00057 &       10.2\\
~GP: SE kernel (5 inputs)  &   0.87 &    0.000 &    0.00093 &    0.00064 &       -7.0\\
Feature space systematics model\\
~GP: SE kernel     &   1.28 &    0.008 &    0.00063 &    0.00051 &       12.9\\
~GP: SSE kernel     &   0.79 &    0.000 &    0.00072 &    0.00047 &      -11.3\\
\hline
\noalign{\smallskip}
\end{tabular}

\end{table*}

\subsection{Stochastic systematics model}
\label{sect:stoc_fits}

Now we have established the reliability of marginalisation over feature space and stochastic processes (GPs) to recover transit parameters in the presence of linear basis models and higher order expansions of them, we might ask whether linear basis models and feature space expansions can reliably recover light curve parameters when the systematics are not exactly described by a simple expansion of basis inputs. This is a more realistic scenario, as systematics are unlikely to be exactly described by such simple functions.

To answer this question, a series of ten thousand light curves were again generated from a basis set $\mathbfss{X}$, only now generating the systematics using a stochastic function of the inputs. For each simulated light curve, two basis functions were generated as before, and a GP with squared exponential kernel was used to generate {\it independent} systematics. This was so cross terms in the feature space expansion did not need to be considered. For each of the two bases in $\mathbfss{X}$, a random draw was taken from a GP with a mean of 1 and squared exponential kernel function:
\[
k(x_{i},x_j) = \exp\left(\frac{(x_i-x_j)^2}{2l^2}\right),
\]
with $l$ equal to 2. This typically resulted in 1-2 turning points per draw, so the generated systematics should be fit reasonably well with low order polynomial expansions of $\mathbfss{X}$. The two draws for each input were then multiplied together, scaled according to $\sigma_r$, and multiplied to the transit model to generate the simulated light curve.

The planet-to-star radius ratio was then recovered as before, using a similar set of linear basis models, marginalisation over feature space expansions of the inputs (to 4th order expansion), Laplacian approximations with shrinkage priors (to 4th order expansion), and GPs using both the SE and SSE kernels. The results are shown in Tab.~\ref{tab:stochastic_results}, and example $\N_\mathrm{s}$ distributions are shown in the bottom panels of Fig.~\ref{fig:histograms}.

The results indicate that marginalisation over feature space is yet again better than model selection, and delivers more reliable results. The marginalisation using the AIC and Laplace approximations provide reliable inference of the transit depth, and the BIC underestimates the uncertainties as for the other systematics models. The most reliable method is now the GP with SSE kernel, which provides almost ideal statistics. The SE kernel also does a good job in recovering the transit depth, despite imposing that the systematics are a joint model of the inputs rather than an independent model. Ideally these tests would be extended to higher order feature spaces and more basis inputs, along with joint systematics models, but this would require too much computation over thousands of light curves.

\begin{table*}
\caption{Results from 10\,000 simulated with the systematics as stochastic functions of basis inputs, as described in Sect.~\ref{sect:stoc_fits}. These were fitted using similar models to those in Tab.~\ref{tab:lbm_results}}
\label{tab:stochastic_results}

\begin{tabular}{lccccc}
\hline
\noalign{\smallskip}
Fitting method & $\chi^2_{\rm r}$ & $\bar{\N_\mathrm{s}}$ & $\bar\sigma_\rho$ & mean acc &  rescaling (\%)\\
\hline
\noalign{\smallskip}
LBM: linear exp       &   6.33 &   -0.000 &    0.00045 &    0.00076 &      151.7\\
LBM: 4th order exp    &   1.36 &   -0.001 &    0.00090 &    0.00078 &       16.6\\
FS: best fit BIC      &   4.60 &   -0.009 &    0.00038 &    0.00057 &      114.4\\
FS: best fit AIC      &   3.31 &   -0.008 &    0.00045 &    0.00060 &       82.0\\
FS: best fit LP       &   3.60 &   -0.014 &    0.00039 &    0.00053 &       89.8\\
FS: marginalised BIC  &   1.78 &   -0.013 &    0.00054 &    0.00052 &       33.4\\
FS: marginalised AIC  &   1.23 &   -0.010 &    0.00065 &    0.00054 &       11.1\\
FS: marginalised LP   &   1.18 &   -0.012 &    0.00067 &    0.00051 &        8.7\\
LP: 4th order exp     &   2.49 &    0.004 &    0.00049 &    0.00056 &       57.8\\
GP: Sqexp/ARD kernel &   1.29 &   -0.038 &    0.00061 &    0.00049 &       13.8\\
GP: Sqexp/SUM kernel &   0.98 &    0.000 &    0.00071 &    0.00051 &       -1.2\\
\hline
\noalign{\smallskip}
\end{tabular}

\end{table*}

\section{Practical implementation}
\label{sect:practical}

These results show that marginalisation over feature space models is the best way to recover light curve parameters in the presence of instrumental systematics described by (independent) low order polynomial expansions of basis inputs. This method can also recover reliable transit parameters when the systematics are not exactly given by a low order polynomial of the basis inputs, but well approximated by it. However, these results were based on simulated light curves, and in practice several other considerations should be taken into account.

To compute the evidence required for marginalising over feature space, both \AICC\ and Laplacian approximations to the evidence are better than the more commonly used BIC. Technically, the Laplacian approximation should give a better estimate of the evidence, especially if the evidence is re-computed from the full distribution once the priors on the weight parameters are fixed, but the \AICC\ is more convenient to calculate. The methods used here also have the added bonus of being calculable directly from a Levenberg-Marquardt optimisation, which could significantly speed up computation time.

These results also show that methods employing model selection with BIC can commonly underestimate the uncertainties by $\sim100\%$. However, this does not mean we should revise the uncertainties of all data extracted using such techniques by this amount, as in practice most authors take more care to select the systematics models and ensure that the model provides an acceptable fit to the data. For example, one could use the residuals from a simple white light curve fit in order to inform the model selection and the required complexity. Furthermore, if the model selection criteria show a clear preference for one model over the others, then model selection and marginalisation should produce identical results, and the choice of model selection criteria will become less important. As the quality of the data increases, model selection will also become easier, as was shown in Sect.~\ref{sect:fs}. This means that for high quality data such as STIS light curves \citep[\eg][]{Sing_2011}, the best fit systematics model might be well constrained, and marginalisation over many models is unnecessary and again might produce the same results.

Transmission and eclipse spectroscopy measurements also have the added bonus of multiple, simultaneous light curves from the same instrument. This should in principle make model selection easier. The current results show that more complex models are a more reliable way to model systematics, suggesting that authors should select the most complex systematics model required for any of the wavelength dependent light curves to achieve more robust inference. Marginalisation over multiple models is still probably a better approach, as we should not expect systematics to behave the same way in all wavelength channels, particularly if the dominant effects are related to pixel-to-pixel sensitivity variations, \ie the systematics change spatially over the detector, which is the case for many instruments. 

Simultaneous observations of multiple light curves also have the advantage of being able to remove common mode systematics. These are systematics that are common to all wavelength channels, and can be determined from analysis of the white light curve. In this case the results are conditioned on the common mode correction, but this is generally not problematic as long as additional systematics that are inserted into the light curves are taken into account during inference. Marginalisation over a common mode correction could be achieved by simultaneously modelling all the light curves, although this is difficult in practice.

When the systematics do not follow a simple polynomial expansion of the basis inputs, then Gaussian processes become the method of choice for modelling instrumental systematics. This is unsurprising, seeing that the feature space expansion does not contain the correct systematics model, and therefore our inference is not always so reliable. In the case of Gaussian processes using a SE kernel (or SSE), we are in effect using an infinite number of basis functions, and can be  sure that the true systematics model, or something very close, is contained within the model space.

The model selection criteria used in this paper also rely on the posterior distribution being well approximated by a Gaussian distribution. This may not always be the case, for example if we fit for all of the transit parameters, or when using other light curve models. One could use more sophisticated methods to calculate the evidence, but care must be taken to ensure that the priors are selected in an objective manner. When the posterior is no longer well approximated by a Gaussian, the Gaussian mixture model used to marginalise over the feature space models may also be inappropriate. This is certainly an advantage for GPs, which do not rely on calculation of the model evidence, and should be applicable to highly non-Gaussian posterior distributions. Arguably, kernels still need to be selected for GPs, which could be performed with model evidence and even marginalisation over many kernels, but the flexibility in even the simplest GP kernels make this less critical.

Nevertheless, marginalisation over a high enough order feature space can still provide reliable results. This could become prohibitively expensive as the order of expansion and the number of basis inputs increases, particularly if one wants to include cross terms in the feature space expansion. Gaussian process models are therefore a better choice of systematics model if $N$ is small enough for GPs to remain computationally feasible. This is typically for $N \lesssim 1000$. Unfortunately for GPs, the computational complexity scales with $\mathcal{O}(N^3)$. This leaves feature space models, with complexity scaling as $\mathcal{O}(N)$, as a more feasible approach for large $N$. They are therefore applicable to larger datasets such as Spitzer light curves without the necessity to bin the light curves, but going to high order feature space will still be challenging.

Another important consideration in using either feature space or GP models, is whether the systematics are well described by a function of the available inputs. For example, the dominant systematic could be a function of a detector property that we have no measurement of (or too noisy a measurement). In this case, the additional systematics could be modelled as a time-dependent function, and could be taken into account by using a stochastic process. For GPs this would require adding a time-dependent component to the kernel, and for feature space models we could use the wavelet method of \citet{Carter_2009} to compute the likelihood for each model.

Finally, it is worth noting that no systematics modelling technique will provide perfect, reliable statistics in all cases. Rather than picking a single technique to extract transit parameters, the best approach is always to use multiple techniques and test whether the inference is dependent on subjective choices made. Where multiple epochs of data exist for the same planetary system, either with the same instrumental setup or with overlapping wavelength regions, common techniques should be tested to validate their use on each instrument.

\section{Conclusions}
\label{sect:con}

This study has investigated the use of common model selection criteria used in the analysis of exoplanet light curves in the presence of instrumental noise, which remains of central importance in our understanding of exoplanet atmospheres. The use of marginalisation over multiple, parametric systematics models was introduced, and demonstrated on a series of transit light curves with known systematics functions. This is shown to be a much more reliable method to recover light curve parameters than simple model selection. Of particular note, the commonly used BIC is shown to be the worst model selection criteria out of those tested, and can result in underestimated uncertainties if care is not taken during inference, therefore low significance claims using such techniques should be treated with caution.

Gaussian process models were also discussed in the context of model selection, and their use was demonstrated on similar sets of simulated transit light curves. They were shown to be an extremely flexible and reliable method to recover transit parameters, even when the kernel used is not the most appropriate. Given the flexibility of GP models, these are perhaps the best method to used when the number of data points is relatively small ($N\lesssim1000$). Feature space models however can also provide reliable inference when the systematics are well approximated by low order polynomials of the input basis set, and are therefore more easily applied to large data sets such as Spitzer light curves.

The technique of marginalising over systematics models contributes to the growing number of statistical tools that we can use to infer parameters from light curves. The choice of systematics model and model selection methods are clearly critical to performing reliable inference. We should therefore take a critical look at our choices of systematics models, and assess how sensitive our results are to subjective decisions. If the results change with subjective decisions, then we should not trust the light curve parameters, or indeed any extraction of atmospheric data based on them. Furthermore, the choice of systematics model as well as the parameters of the systematics model will always contribute to the error budget; therefore pushing to the photon limit is not feasible in the presence of significant systematics. When in doubt, an extra `layer' of Bayesian inference can be used to marginalise over our ignorance of the systematics to reflect our lack of understanding, and allows us to perform more objective inference. Re-analysing and re-observing targets, in particular using different instruments, are invaluable in assessing the validity of the most commonly used systematics models, and therefore in re-evaluating what we truly know about exoplanet atmospheres.

\section*{Acknowledgments}
I am extremely grateful to Suzanne Aigrain and the anonymous referee for a careful reading of the manuscript, and for insightful comments that significantly improved the clarity of the paper.

%

\bibliography{../MyBibliography} 
\bibliographystyle{mn2e} 

\label{lastpage}

\end{document}